**Reply to Comment by D. Spemann *et al*** [*EPL* **98** (2012) 57006]

M. Sepioni, R. R. Nair, I.-Ling Tsai, A. K. Geim and I. V. Grigorieva

School of Physics & Astronomy, University of Manchester, Manchester M13 9PL, UK

Graphite exhibits complex electronic and magnetic properties, and some of the observed anomalies were occasionally interpreted as signs of integer and fractional quantum Hall effects, a field-induced metal-insulator transition, high-temperature ferromagnetism, a large semiconducting gap and even room-temperature superconductivity. Some of the authors of the comment (P. Esquinazi, Y. Kopelevich) are behind all these claims that, for weak evidence, have been met with understandable skepticism.

One of graphite's anomalies stands out, however, being reported by many groups over the last two decades. It is a magnetization signal in samples of highly oriented pyrolytic graphite (HOPG), which indicates that a tiny portion ($\sim 10^{-5}$) of their volume is ferromagnetic at room temperature and above. This ferromagnetism presents a serious puzzle, and no convincing theoretical explanation has emerged so far (see, e.g., recent papers [1,2]). The uncertainty has led to a widely-spread sentiment that the observed signals are due to contamination (see, e.g., [3,4]) as opposed to a positive-case scenario that grain boundaries and edges are ferromagnetic [5-7].

During our investigation of graphene's magnetic properties [8,9] we observed the same ferromagnetic loops in our starting HOPG material. No surprise. Trying to clarify their origin, we found that widely used grades of HOPG were contaminated with micron-size magnetic particles located mostly at grain boundaries. The amount of the detected magnetic material was sufficient to explain the observed signals. Moreover, we found a clear correlation between the amount of magnetic particles for a particular sample and the measured ferromagnetic signal; some HOPG crystals (SPI grade) showed no magnetic contamination and no ferromagnetism, either. Accordingly, our report [10] warned about the contamination and offered a practical method of visualizing the particulate by widely-available backscattering scanning electron microscopy.

The relevance of our work is exemplified perhaps by the results of ref. [7]. This particular paper specifies the used HOPG as ZYH from NT-MTD, one of the grades that we found to be contaminated. We speculated that the observed magnetic particles may explain the reported room-temperature ferromagnetism at grain boundaries [7]. Obviously, it is up to the authors who used such contaminated grades (e.g., [7,11,12]) and those who did not specify their source material to check on our suggestion and either keep or change their conclusions.

Spemann et al argue that their specific HOPG samples were uncontaminated, being checked by the Particle Induced X-ray Emission (PIXE). As just said, fair enough. However, we have to note that we cited their early study [5] because Spemann et al used pristine HOPG of grades similar to those reported by us. Then they argued that the observed ferromagnetic signals were intrinsic and could not be explained by the detected magnetic impurities (several µg/g as in ZY-grades) because "the Fe concentration measured in the HOPG samples was found [by using PIXE] to be homogenously distributed". Several years later [13], they used laterally resolved PIXE to find out that this assumption was incorrect, and Fe impurities in fact formed clusters, large enough to exhibit the conventional ferromagnetism. If Spemann et al accepted that the ferromagnetic signals in those HOPG samples could be explained by the magnetic particulate, as the combination of the two reports implies, this acknowledgment would be helpful for the community. Then, other researchers could have avoided the use of the contaminated starting material as, for example, in refs [7,11,12].

Instead, the comment by Spemann et al makes the subject of intrinsic ferromagnetism in pristine HOPG even murkier by throwing into the mix a completely separate issue of ferromagnetism in irradiated HOPG. We reported on pristine HOPG only, as clearly stated in our abstract, conclusions and throughout the text. We did not mention irradiated or defected graphite, nano-graphite, etc. (e.g., refs. [14,15]). It is not "whatever", as Spemann et al write.

To make our position crystal clear, in a related paper [9], we have studied magnetic properties of irradiated and defected graphene. We found strong paramagnetism, in agreement with theory, but no sign of ferromagnetism even at liquid-helium temperatures. By no means we consider this as evidence against ferromagnetism in all types of graphitic samples and, especially, at the edges (e.g., [16]). If paramagnetic defects are spaced even closer than achieved in ref. [9], their interaction may in our opinion turn paramagnetism into ferromagnetism at accessible temperatures.

We hoped that our report [10] clarified the issue of possible artifacts in pristine HOPG and researchers would finally stop using a contaminated starting material. Unfortunately, the comment has brought the subject back by claiming that other pristine HOPG samples show the same level of ferromagnetism but without a magnetic particulate. We challenge Spemann et al to provide such samples so that we can check their claim independently and report to EPL. We remain open minded but, for the moment, our opinion is based on the fact that all our ZY-grade samples showed both commonly-reported ferromagnetism and a magnetic particulate whereas SPI-grade HOPG showed little magnetic contamination and no ferromagnetism either [10].